\documentclass[a4wide]{article}

\usepackage{epsfig}
\usepackage{amsfonts}

\newcommand{\lb}[1]{\label{#1}}

\newcommand{\bc}{\begin{center}}
\newcommand{\ec}{\end{center}}
\newcommand{\be}{\begin{equation}}
\newcommand{\ee}{\end{equation}}
\newcommand{\bea}{\begin{eqnarray}}
\newcommand{\eea}{\end{eqnarray}}
\newcommand{\ba}[1]{\begin{array}{#1}}
\newcommand{\ea}{\end{array}}

\newcommand{\bt}[1]{\begin{table}[ht]\centering\begin{tabular}{#1}}
\newcommand{\et}[1]{\end{tabular}\caption{\small#1}\end{table}}
\newcommand{\fig}[3]{\begin{figure}[htb]\epsfxsize=80mm\bigskip\centerline{\epsfbox{#1}}\caption{\small\it #2 \label{#3}}\bigskip\end{figure}}

\title{$U_e(1)\times U_g(1)$ Actions in $2+1$--Dimensions:\\
Full Vectorial Electric and Magnetic Fields}

\author{P. Castelo Ferreira\\ \small CENTRA, Instituto Superior T\'ecnico, Av. Rovisco Pais, 1049-001 Lisboa, Portugal\\ {\tt pcastelo@catastropha.org}}

\begin{document}

\maketitle

\abstract{It is considered a dimensional reduction of $U_e(1)\times U_g(1)$
$3+1$--dimensional electromagnetism with a gauge field (photon)
and a pseudo-vector gauge field (pseudo-photon) to $2+1$--dimensions.
In the absence of boundary effects, the quantum structure is maintained, while
when boundary effects are considered, as have been previously
studied, a cross Chern-Simons term between both gauge fields is present, which
accounts for topological effects and changes the quantum structure of the theory.
Our construction maintains the dimensional reduced action invariant under
parity ($P$) and time-inversion ($T$). We show that the theory has
two massive degrees of freedom, corresponding to the longitudinal modes
of the photon and of the pseudo-photon and briefly discuss the quantization
procedures of the theory in the topological limit (wave functional
quantization) and perturbative limit (an effective dynamical
current theory), pointing out directions to solve the constraints
and deal with the negative energy contributions from pseudo-photons.
We recall that the physical interpretation of the fields in the planar
system is new and is only meaningful in the context of $U_e(1)\times U_g(1)$
electromagnetism. In this work it is shown that all the six electromagnetic vectorial fields
components are present in the dimensional reduced theory and that,
independently of the embedding of the planar system, can be described
in terms of the two gauge fields only. As far as the author is aware
it is the first time that such a construction is fully justified,
thus allowing a full vectorial treatment at variational level of
electromagnetism in planar systems.\\[5mm]
PACS: 03.50.De, 41.20.-q, 13.40.–f\\
Keywords: Electromagnetism, Two Dimensions, pseudo-photon}

\section{Introduction}
Several physical phenomena are effectively two dimensional, for example Bloch electrons~\cite{Bloch}, superconductivity~\cite{SC}
and the Hall effect~\cite{HE_1}, to name a few. Describing electromagnetism at variational level in $2+1$--dimensions
lead us to the respective Maxwell action and electromagnetic field definitions
\be 
\ba{l}
\displaystyle S_{\textrm{Maxwell}}=\int dx^3 F_{\mu\nu}F^{\mu\nu}\ ,\\[5mm]
\displaystyle E^i=F^{0i}\ (i=1,2)\ ,\ \  B=F_{12}\ ,
\lb{S_M}
\ea
\ee
such that in relation to $3+1$--dimensional electromagnetism, only the planar electric field components and the
orthogonal magnetic field component are present, and the theory has no propagating degrees of freedom.
For many models and theories involving electromagnetic fields in planar or embedded systems these are
enough to describe the physics in question. Also, for some application where boundary effects are relevant,
topological effects are considered through the inclusion of a topological Chern-Simons term
\be
S_{\textrm{CS}}=m\int dx^3 \epsilon^{\mu\nu\lambda}A_{\mu}F_{\nu\lambda}\ .
\lb{S_CS}
\ee
This term is of lower order than the Maxwell term (it has only one derivative in the fields) and in some applications
it is dominant in relation to the Maxwell term. It changes the quantum structure of the theory and gives a topological
mass $m$ to the gauge field $A$~\cite{mass} which implies that in the perturbative regime the theory has one propagating
degrees of freedom (longitudinal massive mode), as opposed to the Maxwell theory. However it is, generally
non-renormalizable~\cite{F2_CS} and explicitly breaks parity $P$ and time-inversion $T$ symmetries.
In addition to having propagating degrees of freedom, it is often justified theoretically, as a quantum
correction to the Maxwell action~\cite{F2_CS}. From a more phenomenological perspective it has
been used as an effective description of the fractional hall effect~\cite{HE_2} through a collective gauge
field and it also describes boundary conformal field theories~\cite{CFT}.
However a relevant question can be asked, can we have a description of planar electromagnetism
containing the full vectorial electromagnetic fields as we do in standard electromagnetism?
Possible approaches to answer this question are to consider four dimensional descriptions of the
planar physics based in extensions of the Maxwell action including Kalb-Ramon form fields~\cite{AG_1} or to consider
dimensional reduction of standard electromagnetism such that the gauge field component orthogonal to the planar
system is described effectively by a scalar field~\cite{dim_red}. The physical meaning of a Kalb-Ramon form field~\cite{KR}
is to effectively describe either dynamical vortexes~\cite{vortexes} or dynamical currents~\cite{currents}.
Usually imply chiral symmetry breaking and can also be described effectively as planar Chern-Simons theories~\cite{Kogan},
which take us back to the original Maxwell Chern-Simons theories already discussed, hence our question remains pertinent.
The scalar field obtained upon dimensional reduction, depending in the boundary conditions and symmetries of the
embedding in the four dimensional system, may be null (this is the case, for example, for standard Neumann boundary
conditions). In what follows we will consider another possible construction by deriving a dimensional
reduction of electromagnetism with one vector gauge field (the standard photon) and one pseudo-vector gauge field
(pseudo-photon). We will obtain directly a boundary cross Chern-Simons term that accounts for topological boundary effects and simultaneously
preserves $P$ and $T$ symmetries, this feature is not new in our work and have been studied in detail in~\cite{Witten,Schwarz}.
Also this theory accommodates the six components of the physical electric and magnetic fields defined in terms of the gauge fields
and independently of the boundary conditions and embedding. This feature is, in principle, physically appealing. Even being
in a planar system, we are still dealing with the same physics, in particular electromagnetism. Also we are considering here a
dimensional reduced theory which is theoretically consistent with the four dimensional Maxwell equations. We note that originally $U_e(1)\times U_g(1)$
has been justified by the inclusion of magnetic monopoles~\cite{CF,action_00,dual,Sing} maintaining the field configurations regular,
i.e. free of extended singularities as the Dirac String and Wu-Yang fiber-bundle~\cite{Dirac}. Although the existence of magnetic
monopoles is still the best justification for electric charge quantization~\cite{Dirac}, so far they have not been experimentally
observed. However we remark that it is enough to consider non-trivial background electric fields ($\nabla.E\neq 0$) or magnetic fields
($\dot{B}\neq 0$) to justify the existence, at functional level, of pseudo-photons~\cite{pseudo_nt} and justify extended $U_e(1)\times U_g(1)$
electromagnetism. There will be new problems associated to this theory that we shortly discuss by the end of this work,
namely due to the pseudo-photon being a ghost (or phantom, it has a negative kinetic term in the Hamiltonian),
hence the quantization of the theory is not a straight forward task. Nevertheless, this apparent nuisance, may turn out to justify the
low energy of the Laughlin's wave functions due to the negative contributions of pseudo-photon excitations, as was put forward
in~\cite{FQHE_AC} using a semi-classical model. Also the model of~\cite{FQHE_AC} proofs the equivalence between Dirac's quantization
condition and the, experimentally verified, quantization of magnetic flux. Moreover a more fundamental description of any theory
underlines a better understanding of the theory, therefore more reliable prediction and control of physical systems.

We consider as starting point the $3+1$--dimensional action for $U_e(1)\times U_g(1)$ electromagnetism introduced in~\cite{action_00,dual}
given by $S_{4}=\int dx^4{\mathcal{L}}_4$ and Lagrangian density
\be
\ba{rcl}
\displaystyle{\mathcal{L}}_4&=&\displaystyle-\frac{1}{4}F_{IJ}F^{IJ}+\frac{1}{4}G_{IJ}G^{IJ}\\[2mm]
               & &\displaystyle+\frac{1}{4}\epsilon^{IJKL}F_{IJ} G_{KL}+A_I\,J_e^I\ ,
\ea
\lb{S_4}
\ee
where the gauge connections are $F_{IJ}=\partial_I A_J-\partial_J A_I$ and $G_{IJ}=\partial_I C_J-\partial_J C_I$ with the space-time indexes
$I=\perp,\mu$  such that $\perp$ stands for the spatial direction orthogonal to the planar system and $\mu=0,1,2$ correspond
to the $2+1$--dimensional space-time indexes.

\section{Dimensional Reduction}

In this section we address a possible dimensional reduction scheme for action~(\ref{S_4}). We consider
a planar system of a certain thickness $\delta_\perp$ with two boundaries $\Sigma_1$
and $\Sigma_2$ as represented in figure~\ref{fig.slab}.
\fig{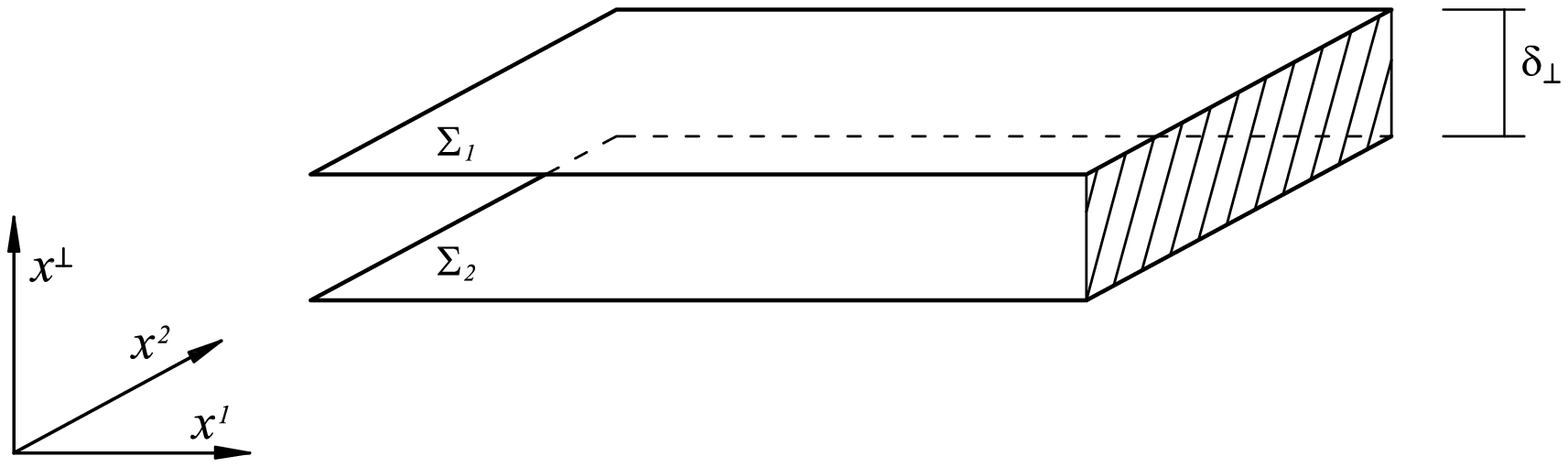}{$2+1$--dimensional system of thickness $\delta_\perp$, with boundaries $\Sigma_1$ and $\Sigma_2$, embedded in a $3+1$--dimensional system.}{fig.slab}
Let us take in consideration the regularity of the each of the gauge fields, such that
the constants $r_A=0,1$ and $r_C=0,1$ are set accordingly to:
\be
r_{A(C)}=0,1:\ A(C)\ \mathrm{is\ regular,non-regular}
\lb{reg}
\ee
Then taking a decomposition of the Maxwell and current terms in action~(\ref{S_4}) into the $2+1$-dimensional components $\mu$ and the
orthogonal direction $x^\perp$ and integrate by parts the Hopf term containing derivatives along $x^\perp$, we obtain
\be
\ba{l}
\displaystyle{\mathcal{L}}_4=-\frac{1}{4}F_{\mu\nu}F^{\mu\nu}+\frac{1}{4}G_{\mu\nu}G^{\mu\nu}+A_\perp\,J_e^\perp+A_\mu\,J_e^\mu\\[3mm]
-\frac{1}{2}\partial_\perp A_\mu\partial^\perp A^\mu+\partial_\perp A_\mu\partial^\mu A^\perp-\frac{1}{2}\partial_\mu A_\perp\partial^\mu A^\perp\\[3mm]
+\frac{1}{2}\partial_\perp C_\mu\partial^\perp C^\mu-\partial_\perp C_\mu\partial^\mu C^\perp+\frac{1}{2}\partial_\mu C_\perp\partial^\mu C^\perp\\[3mm]
\displaystyle+\epsilon^{\perp\mu\nu\lambda}\partial_\perp\left(A_{\mu}\partial_{\nu}C_{\lambda}\right)+\epsilon^{\mu\nu\perp\lambda}\partial_\perp\left(\partial_{\nu}A_{\mu}C_{\lambda}\right)\\[3mm]
\displaystyle-r_C\,\epsilon^{\perp\mu\nu\lambda}A_{\mu}\partial_\perp\partial_{\nu}C_{\lambda}-r_A\,\epsilon^{\mu\nu\perp\lambda}\left(\partial_\perp\partial_{\nu}A_{\mu}\right)C_{\lambda}\\[3mm]
               \displaystyle+r_Cr_A\,\epsilon^{\mu\perp\nu\lambda}\partial_\nu A_{\perp}\partial_{\mu}C_{\lambda}+r_Cr_A\,\epsilon^{\mu\nu\lambda\perp}\partial_\nu A_{\mu}\partial_{\lambda}C_{\perp}.
\ea
\lb{L_decomp}
\ee
We are also taking the following assumptions:
\begin{enumerate}
\item The fields are localized in the range
\be
x^\perp\in\left[-\delta_\perp/2,+\delta_\perp/2\right]\ ,
\ee
and are slowly varying over the orthogonal direction such that the integral over $x^\perp$
can be performed in the above range.
\item Gauge transformations are fixed along the orthogonal direction, such that
\be
\partial_\perp\Lambda=0\ .
\lb{gauge}
\ee
\item The orthogonal derivatives of the fields constitute boundary conditions of the system such that their effects
are manifested at the level of the action through the \textit{external} currents
\be
J_A^\mu=\partial^\perp A^\mu\ \ ,\ \ \ J_C^\mu=\partial^\perp C^\mu\ .
\lb{perp_cond}
\ee
\item The field components $A_\perp$ and $C_\perp$ are, in the $2+1$--dimensional system, not charged under any of the
$U(1)$'s and are identified to scalar fields~\cite{dim_red}
\be
\phi=A_\perp\ \ ,\ \ \ \ \varphi=C_\perp\ .
\ee
\item We are taking in account the boundary terms due to the integration by parts of the Hopf term
(here we are referring to the last terms in equation~(\ref{L_decomp})). In order to do so we
identify $\Sigma(x^\perp=0)\cong\Sigma_1\cong\Sigma_2$ and considering the boundary action~\cite{Witten}
\be
\int_{\Sigma_1-\Sigma_2}\epsilon^{\mu\nu\lambda}A_\mu\partial_\nu C_\lambda\equiv \frac{k}{2}\,\int_\Sigma\epsilon^{\mu\nu\lambda}A_\mu\partial_\nu C_\lambda\ ,
\lb{id}
\ee
where the value of $k=+1,0,-1$ depends on the specifications of the system, in particular of the embedding and symmetries of the $2+1$--dimensional
system in relation to the $3+1$--dimensional system~\cite{Witten}.
\end{enumerate}
We can now perform the integration over the orthogonal coordinate $\int_{-\delta_\perp/2}^{+\delta_\perp/2} dx^\perp=\delta_\perp$
obtaining the dimensional reduced action $S_{3}=\delta_\perp\int dx^3{\mathcal{L}}_3$ with Lagrangian density
\be
\ba{rl}
{\mathcal{L}}_3=&\displaystyle-\frac{1}{4}F_{\mu\nu}F^{\mu\nu}+\frac{1}{4}G_{\mu\nu}G^{\mu\nu}+\frac{1}{2}J_A^\mu J_{A\,\mu}\\[3mm]
                &\displaystyle-\frac{1}{2}\partial_\mu\phi\partial^\mu\phi+\frac{1}{2}\partial_\mu\varphi\partial^\mu\varphi-\frac{1}{2}J_C^\mu J_{C\,\mu}\\[3mm]
                &\displaystyle-\frac{r_C}{2}\epsilon^{\mu\nu\lambda}F_{\mu\nu}J_{C\,\lambda}+\frac{r_A}{2}\epsilon^{\mu\nu\lambda}G_{\mu\nu}J_{A\,\lambda}\\[3mm]
                &\displaystyle-\frac{r_Ar_C}{2}\epsilon^{\mu\nu\lambda}\partial_\mu\phi G_{\nu\lambda}+\frac{r_Ar_C}{2}\epsilon^{\mu\nu\lambda}\partial_\mu\varphi F_{\nu\lambda}\\[3mm]
                &\displaystyle+\partial_\mu\phi\,J^\mu_A-\partial_\mu\varphi\,J^\mu_C+\phi\,J_e^\perp+A_\mu\,J_e^\mu\\[3mm]
                &\displaystyle+\frac{k}{4\delta_\perp}\,\epsilon^{\mu\nu\lambda}A_\mu\partial_\nu C_\lambda+\frac{k}{4\delta_\perp}\,\epsilon^{\mu\nu\lambda}C_\mu\partial_\nu A_\lambda.\\[3mm]
\ea
\lb{S_3}
\ee
We have perform an integration by parts in writing the current terms for $J_A^\mu$ and $J_C^\mu$ and the terms containing the
$2+1$--dimensional Levi-Civita symbol swap sign due to the relation $\epsilon^{\perp\mu\nu\lambda}=-\epsilon^{\mu\nu\lambda}$.
We note that the square terms in the currents $J_A^\mu J_{A\,\mu}=(\partial^\perp A^\mu)^2$ and $J_C^\mu J_{C\,\mu}=(\partial^\perp C^\mu)^2$
are non-dynamical constants which stand for boundary conditions as given by~(\ref{perp_cond}). These are relevant only as a contribution to
the vacuum energy being relevant, for example, when one considers gravity~\cite{lambda}, such that they play the role of an
effective $2+1$--dimensional cosmological constant $4\lambda^{\mathrm{eff}}=(\partial^\perp A^\mu)^2-(\partial^\perp C^\mu)^2$.
In this work these terms play no role and we may simply consider that its effect corresponds to a vacuum energy shift.
We further note that our $2+1$--dimensional action is, as it stands, gauge invariant as long as $\phi$ and $\varphi$
are not charged under any of the gauge groups $U(1)$'s. We recall that we gauge fixed along the orthogonal
direction~(\ref{gauge}) and that the coupling between the scalar fields $\phi$ and $\varphi$ and the
gauge fields $A$ and $C$ is done only through the currents $J_A$ and $J_C$. Extensions of this construction can be implemented such that
the scalar fields are charged under the gauge groups, in such case it is necessary to consider gauge covariant derivatives
such that the action remains gauge invariant~\cite{mass_1}. We do not develop this topic here. We also remark that the middle terms 
in the second and third line of equation~(\ref{L_decomp}) could be integrated by parts, however this would hold boundary terms
that are not gauge invariant.

\section{Electromagnetic Fields, Canonical Momenta and Boundary Conditions}

From the definitions of electric and magnetic field in the $3+1$--dimensional system~\cite{dual,Sing}
we obtain the physical electric and magnetic fields definitions in the
planar system:
\be
\ba{rcl}
E^\perp&=&\displaystyle\partial^0\phi-J^0_A-G_{12}\ ,\\[3mm]
E^i&=&\displaystyle F^{0i}-\frac{r_C}{2}\epsilon^{ij}\left(r_A\partial_j\varphi-J_{C\,j}\right)\ ,\\[3mm]
B^\perp&=&\displaystyle-\partial^0\varphi+J^0_C+F_{12}\ ,\\[3mm]
B^i&=&\displaystyle-G^{0i}+\frac{r_A}{2}\epsilon^{ij}\left(r_C\partial_j\phi-J_{A\,j}\right)\ ,
\ea
\lb{fields}
\ee
where we have taken in consideration the regularity of each of the gauge fields as given by~(\ref{reg}).
Concerning the canonical momenta for our action we obtain six independent ones
\be
\ba{rcl}
\pi_A^i&=&\displaystyle-F^{0i}+\frac{r_Ar_C}{2}\epsilon^{ij}\partial_j\varphi-\frac{r_C}{2}\epsilon^{ij}J_{C\,j}+\frac{k}{4\delta_\perp}\epsilon^{ij}C_j,\\[3mm]
\pi_C^i&=&\displaystyle+G^{0i}-\frac{r_Ar_C}{2}\epsilon^{ij}\partial_j\phi+\frac{r_A}{2}\epsilon^{ij}J_{A\,j}+\frac{k}{4\delta_\perp}\epsilon^{ij}A_j,\\[3mm]
\pi_\phi&=&\displaystyle-\partial^0\phi+J^0_A-r_Ar_C\,G_{12},\\[3mm]
\pi_\varphi&=&\displaystyle+\partial^0\varphi-J^0_C+r_Ar_C\,F_{12}.
\ea
\lb{pi}
\ee
We note that the relations between the canonical conjugate momenta to $A_i$ and $C_i$~(\ref{pi}) and the electric and
magnetic fields~(\ref{fields}) hold a new correction due to the Chern-Simons boundary contribution in relation
to the $3+1$--dimensional relations obtained in~\cite{dual}. Here we are referring to the last terms depending on $k$ in
the definitions of $\pi^i_A$ and $\pi^i_C$ in~(\ref{pi})). This is a common feature of Chern-Simons theories~\cite{mass}.
For completeness on the discussion of definition of canonical momenta, it is relevant to point out that, considering directly the
$2+1$--dimensional theory without any assumptions of the embedding in $3+1$--dimensions, may raise an indefinition of the canonical
structure. This is due to the Chern-Simons term to be defined only up to a boundary term, specifically considering an integration by parts
we have that $\int [A\wedge dC+C\wedge dA]=\int [(1+\xi)A\wedge dC+(1-\xi)C\wedge dA]+\xi\oint A\wedge C$.
This result is already present in the original work by Schwarz~\cite{Schwarz} and 
can be traced back to the choice of a quantization polarization for pure
Chern-Simons theories~\cite{BN}. In simple terms which fields are the canonical coordinates
and which fields are the canonical momenta. When we consider both the Maxwell and Chern-Simons terms this
indefinition can be solved, in~\cite{CKT} it was shown that when all topological
effects are taking in consideration, including the boundary effects due to integration
by parts, the Chern-Simons action for several gauge fields is reduced
to the form $S_{CS}=\int G_{IJ}A^I\wedge dA^J+\oint K_{IJ}A^I\wedge A^J$
with $G_{IJ}$ a symmetric matrix and $K_{IJ}$ a anti-symmetric matrix. In addition it has recently been showed
that, independently of boundary effects, a consistent definition of canonical structure is still possible~\cite{Govaerts}.
To finalize let us stress that, independently of the above discussion, the canonical momenta $\pi^i_A$ and $\pi^i_C$ as
defined in~(\ref{pi}) are directly derived from the $3+1$--dimensional theory, thus being consistent with the higher
dimensional canonical momenta as derived in~\cite{dual}. This is enough to justify the above choice.

So far we have not imposed any particular boundary conditions in our construction. To properly
define the field content of the dimensional reduced theory and its embedding into the 
higher dimensional world it is necessary to do so. The standard types of boundary conditions
correspond usually to gauging symmetries of the fields with respect to the orthogonal
direction $x^\perp$~\cite{Jackson}. Specifically when the orthogonal fields swap sign with respect
to both sides of the planar system we have Neumann boundary conditions, such that they must vanish in
the planar system, i.e. $A_\perp=C_\perp=0$. These conditions correspond to an orbifold under the symmetry
${\mathbb{Z}}_2: x^\perp\to-x^\perp$, usually are applicable to the internal fields (meaning the
fields that are not imposed externally), and the factor of $1/2$ in the Chern-Simons boundary contributions to
the action in~(\ref{id}) can also be justified by this kind of orbifolds~\cite{Horawa}. When the fields
do not change across the orthogonal direction to the planar system we have Dirichlet boundary conditions such that
the orthogonal derivatives of fields vanish in the planar system, i.e. $\partial_\perp A=\partial_\perp C=0$.
This kind of boundary conditions usually applies to the external applied fields. For Neumann boundary conditions for
all fields, generally we have $k\neq 0$, while for Dirichlet boundary conditions for all fields we always
have $k=0$. This is shown by noting that the integrations by parts in the action decomposition~(\ref{L_decomp})
are null for $\partial_\perp A_\mu=0$. This is resumed as
\be
\ba{llcl}
\mathrm{Neumann:}  &A_\perp=C_\perp=0&\Leftrightarrow&\phi=\varphi=0\ ,\\[3mm]
\mathrm{Dirichlet:}&\partial_\perp A_\mu=\partial_\perp C_\mu=0&\Leftrightarrow&J_A^\mu=J_C^\mu=0\ .
\ea
\lb{bc}
\ee
However, generally, any boundary conditions or combination of boundary conditions can be considered.
In particular for systems where spin polarization effects are not relevant (i.e. considering the
magnetic momenta of fermions to be null) and when considering both external and internal fields
we should, consistently with the above discussion, consider Neumann boundary conditions
for the internal fields and Dirichlet boundary conditions for the external fields. Finally considering Neumann boundary
conditions ($\varphi=\phi=0$), regular fields ($r_A=r_C=0$) and constant orthogonal fields
($J_A=\partial^\perp A=0$ and $J_C=\partial^\perp C=0$), we obtain the action, electromagnetic field and canonical momenta definitions
\be
\ba{l}
\displaystyle S_3=\delta_\perp\,\int d^3x\left[-\frac{1}{4}F_{\mu\nu}F^{\mu\nu}+\frac{1}{4}G_{\mu\nu}G^{\mu\nu}\right.\\[3mm]
\ \ \ \ \ \ \displaystyle\left.+\frac{1}{8\delta_\perp}\epsilon^{\mu\nu\lambda}A_{\mu}G_{\nu\lambda}+\frac{1}{8\delta_\perp}\epsilon^{\mu\nu\lambda}C_{\mu}F_{\nu\lambda}\right],\\[5mm]
\displaystyle E^\perp=-G_{12},\ E^i=F^{0i},\ B^\perp=F_{12},\ B^i=-G^{0i},\\[5mm]
\displaystyle \pi_A^i=-F^{0i}+\frac{k}{4\delta_\perp}\epsilon^{ij}C_j,\ \pi_C^i=G^{0i}+\frac{k}{4\delta_\perp}\epsilon^{ij}A_j.
\ea
\lb{S_3_fields}
\ee
Hence we have available the six components of the electric and magnetic
fields as we do in the original $3+1$--dimensional system and its definitions are
fully consistent in the framework of extend $U_e(1)\times U_g(1)$ electromagnetism.
This is a novel feature of our construction which is certainly useful in the description
of planar systems where longitudinal magnetic fields and orthogonal electric
fields are present.

\section{Degrees of Freedom and Quantization}

We have pointed out that one of the motivations of considering the usual Chern-Simons action~(\ref{S_CS}),
is to have propagating degrees of freedom for the photon. Hence it is relevant to ask which are the
propagating degrees of freedom for the action~(\ref{S_3_fields}). Computing the equations
of motion for $A$ and $C$ we obtain respectively
\be
\ba{rcl}
\displaystyle\partial_\mu F^{\mu\nu}+\frac{k}{4\delta_\perp}\epsilon^{\nu\mu\lambda}G_{\mu\lambda}&=&0\ ,\\[3mm]
\displaystyle\partial_\mu G^{\mu\nu}-\frac{k}{4\delta_\perp}\epsilon^{\nu\mu\lambda}F_{\mu\lambda}&=&0\ .
\ea
\lb{eom}
\ee
These equations can be decoupled. By considering the solutions of the above equations
for $G_{\mu\nu}$ and $F_{\mu\nu}$, respectively, and replacing it in the remaining equation,
we obtain
\be
\ba{rcl}
\displaystyle\partial_\mu\partial^\mu A^{\nu}+2\,\left(\frac{k}{2\delta_\perp}\right)^2\,A^{\nu}&=&0\ ,\\[3mm]
\displaystyle\partial_\mu\partial^\mu C^{\nu}+2\,\left(\frac{k}{2\delta_\perp}\right)^2\,C^{\nu}&=&0\ .
\ea
\lb{eom_diag}
\ee
In deriving this result we have used the identity $\epsilon_{\mu\nu\lambda}\epsilon^{\mu'\nu'\lambda}=-2(\delta_\mu^{\ \mu'}\delta_\nu^{\ \nu'}-\delta_\mu^{\ \nu'}\delta_\nu^{\ \mu'})$
and performed an integration such that this result is valid up to a total divergence $\partial^\nu\Lambda$ which
can be offset by an appropriate gauge transformation (or by considering an appropriate gauge choice).
This result shows that we have massive propagating degrees of freedom (one longitudinal mode for the the photon and
another one for the pseudo-photon) with mass
\be
m_A=m_C=\sqrt{2}\frac{|k|}{2\delta_\perp}\ .
\ee
This is actually welcome, we manage to obtain propagating degrees of freedom maintaining $P$ and $T$ invariance.

Concerning quantization of the theory we are faced with new problems that we briefly discuss next, pointing
directions for further developments. From the action and the definitions of canonical momenta given
in~(\ref{S_3_fields}) we obtain the Hamiltonian and Gauss' laws constraints
\be
\ba{rcl}
\displaystyle{\mathcal{H}}&=&\displaystyle+\frac{1}{2}\left(\pi_A^i-\frac{k}{4\delta_\perp}\epsilon^{ij}\partial_i C_j\right)^2+\frac{1}{4}F_{ij}F^{ij}\\[3mm]
& &\displaystyle-\frac{1}{2}\left(\pi_C^i-\frac{k}{4\delta_\perp}\epsilon^{ij'}\partial_i C_{j'}\right)^2-\frac{1}{4}G_{ij}G^{ij}\ ,\\[5mm]
\displaystyle{\mathcal{G}}_A&=&\displaystyle\partial_i\left[\pi_A^i+\frac{k}{4\delta_\perp}\epsilon^{ij}\partial_i C_j\right]\ ,\\[5mm]
\displaystyle{\mathcal{G}}_C&=&\displaystyle\partial_i\left[\pi_C^i+\frac{k}{4\delta_\perp}\epsilon^{ij}\partial_i A_j\right]\ .
\ea
\lb{constraints}
\ee
The Hamiltonian constraint contains two distinct sectors corresponding to the usual photon (the $A$ field) and
the pseudo-photon (the $C$ field). We readily conclude that, as already expected from the $3+1$--dimensional
theory~\cite{dual}, excitations of photons contribute positively to the energy of the quantum state, while
pseudo-photon excitations contribute negatively to the energy of the quantum state. In principle existence of
negative energy states are unwelcome since they violate causality. The most straight forward way to solve
this problem, although not very elegant, is to postulate positive energy for the quantum state such that
the contribution due to excitations of pseudo-photons can never overcome the contribution due to excitations
of photons. However we remark that, in addition to the Hamiltonian constraint, consistently with the
equations of motion~(\ref{eom}), both sectors are related trough the Gauss' laws and,
upon quantization, these constraints must also to be taken in consideration. Depending on the way we implement
these constraints we can have independent excitations, corresponding to each sector, or have a combination
of the degrees of freedom, such that physical excitations correspond to a combination of photon and
pseudo-photon excitations. These two possibilities correspond to the topological regime of the
theory (low-energy) and the perturbative regime of the theory (high-energy). In the topological limit
the  appropriate approach is to consider a functional quantization formalism~\cite{BN,AFC}
such that the quantum constraints~(\ref{constraints}) are solved at the level of wave functionals.
The ground state solution is known to be
\be
\Phi[A,C]=\exp\left\{+i\,\frac{k}{4\delta_\perp}\,\int d^2x\epsilon^{ij}A_iC_j\right\}\ ,
\lb{Phi_AC}
\ee
which leaves both sectors of the theory unconstrain. Excited topological states for Maxwell Chern-Simons
theories have been introduced in~\cite{Dbranes}, however have not been properly consider in the present context,
for which are expected to correspond to electric and magnetic vortexes as put forward in~\cite{FQHE_AC}.
For this case the external fields are expected to drive (control) the excitations of the system such that imposing
energy positiveness is, in principle, not necessary. As for the perturbative limit we can consider the usual
quantum field theory harmonic operator expansion of the fields $A$ and $C$. Then the Gauss' laws will
impose constraints between the harmonic operators, the polarizations and the energies of each field,
such that the degrees of freedom are no longer independent. As we already pointed out this construction can be traced back to the equations
of motion~(\ref{eom}). In order to have some physical insight for this construction, let us solve the equations of motion
for $G$, obtaining directly an effective theory for the photon field only. The Lagrangian and
equations of motion for the effective theory are
\be
\ba{rcl}
\displaystyle{\mathcal{L}}_A&=&\displaystyle-\frac{\delta_\perp^2}{2k^2}\partial_\mu F^{\mu\nu}\partial_{\mu'}F^{\mu'}_{\ \,\nu}+\frac{1}{4}F_{\mu\nu}F^{\mu\nu}\ ,\\[3mm]
0&=&\displaystyle-\frac{2\delta_\perp^2}{k^2}\partial_\mu\partial^\mu\left(\partial_{\mu'} F^{\mu'\nu}\right)-\left(\partial_{\mu'} F^{\mu'\nu}\right)\ .
\ea
\lb{L_A}
\ee
This is an interesting result. We obtain a non-linear massive theory which holds an effective description
in terms of dynamical massive electric currents $j^\nu=\partial_{\mu'}F^{\mu'\nu}=(\partial_iE^i,-\dot{E}^i+\epsilon^{ij}\partial_jB)$.
It has no negative energy excitations, the Maxwell term is playing the role of mass, while the non-linear correction
plays the role of the kinetic term. We cannot avoid to notice some similarities between this
effective non-chiral theory and the four dimensional chiral dynamical current theories~\cite{currents}.
Also we note that, upon inclusion of fermion effects, chiral symmetry breaking may be expected.
We have preliminary indicated the possibilities concerning quantization of the
limiting cases for the theory that allow to avoid strictly negative energy quantum states.
We will develop these issues in detail somewhere else~\cite{progress}.

\section{Conclusions}
We derived a dimensional reduced action for $U_e(1)\times U_g(1)$ electromagnetism containing a vector gauge field (photon)
and a pseudo-vector gauge field (pseudo-photon). We obtain a $2+1$--dimensional action that is $P$ and $T$ invariant,
has massive propagating degrees of freedom and, independently of the boundary conditions and embedding of the planar system,
accounts for the full six vectorial components of the electromagnetic fields in terms of the gauge fields only.
Gauge invariance and Lorentz invariance are, as expected, reduced down to $2+1$--dimensions only.
As far as the author is aware it is the first time that such a construction is achieved, being
only fully justified in the framework of $U_e(1)\times U_g(1)$ electromagnetism, and allowing
a full vectorial treatment, at variational level, of electromagnetism in planar systems.
We have also discuss the quantization procedures which allow to avoid strictly negative
quantum states both the topological and perturbative regimes. A full study will follow
somewhere else~\cite{progress}. As an example of the relevance of the construction presented
here it is applied to the fractional Hall effect in~\cite{FQHE_AC}. Other practical experimental
frameworks where our construction may be relevant is, for example, topological superconductivity
in Josephson junctions~\cite{AG_1,AG_2}, BEC phase transitions in superconductors~\cite{Rivers}
and in bi-layer Hall systems~\cite{bilayer}.\\[5mm]

\noindent{\bf Acknowledgments}\\
The author thanks the referees for valuable suggestions and comments that significantly improved the manuscript.
This work is supported by SFRH/BPD/17683/2004.

\end{document}